\documentclass[a4paper,11pt]{article}
\usepackage{pos}

\title{Light Cone Distribution Amplitudes of Heavy-Light Mesons and Quarkonia}

\author*[a,b]{Fernando E. Serna}
\author[a]{Bruno El-Bennich}

\affiliation[a]{Laboratorio de F\'isica Te\'orica e Computacional,  Universidade Cidade de S\~ao Paulo\\
 Rua Galv\~ao Bueno 868, 01506-000 S\~ao Paulo, SP, Brazil.}

\affiliation[b]{Departamento de F\'isica, Universidad de Sucre, \\
 Carrera 28 No. 5-267, Barrio Puerta Roja, Sincelejo, Col\'ombia}

\emailAdd{fernando.enrique@unesp.br,  bruno.bennich@cruzeirodosul.edu.br}

\abstract{In this work we use the framework of the Dyson-Schwinger and Bethe-Salpeter equations to compute Light-Cone Distribution Amplitudes of heavy-light mesons and quarkonia. In studying the meson
 properties, we  introduce a flavor dependence in the heavy-quark sector of  the Bethe-Salpeter ladder kernel which yields improved numerical results for masses and leptonic decay constants of the pseudoscalar 
 $D$, $D_s$, $B$ and $B_s$ mesons.  Finally, the corresponding heavy-light Bethe-Salpeter amplitudes are projected onto the light front and we reconstruct the distribution amplitudes of the mesons in the full theory.} 

\FullConference{%
  *** 10th International Workshop on Charm Physics (CHARM2020), ***\\
  *** 31 May - 4 June, 2021 ***\\
  *** Mexico City, Mexico - Online ***
}

\begin{document}
\maketitle


\section{Introduction}

The introduction of hadronic light-cone distribution amplitudes (LCDA) dates back to the seminal works on hard exclusive reactions in perturbative 
QCD~\cite{Chernyak:1977fk,Efremov:1978rn,Efremov:1979qk,Lepage:1979zb,Lepage:1980fj}. These nonperturbative and scale-dependent functions can be understood as 
the closest relative of quantum mechanical wave functions in quantum field theory. They describe the longitudinal momentum distribution of valence quarks in a hadron in the 
limit of negligible transverse momentum, here given by the leading Fock-state contribution to its light-front wave function, the so-called leading-twist LCDA. In particular, 
the light-front formulation of a wave function allows for a probability interpretation of partons not readily accessible in an infinite-body field theory, since particle number 
is conserved in this frame. In other words,  $\phi(x,\mu)$ expresses the light-front fraction of the hadron's momentum carried by a valence quark. 

In this work, we focus on the study of the LCDAs of heavy-light systems. The interest in these light-front wave functions stem from decade-old calculations in QCD factorization (QCDF), which only employed asymptotic distributions of the form $\phi(x)\approx x(x-1)$ or simple model distributions. The motivation for these calculations is due to the desire to reduce theoretical uncertainties, so as to make more reliable predictions for CP violation (matter-antimatter asymmetry) and the Cabibbo-Kobayashi-Maskawa parameters (CKM) which could indicate the presence of new physics~\cite{Charles:2004jd}.

This is because in weak heavy-meson decays, such as non-leptonic $D$ decays into two lighter mesons, direct CP violation originates in the interference of at least two amplitudes with different weak and strong phases. The weak phase, determined by combining CKM matrix components, is extracted from a CP violating asymmetry. The strong phase has to be calculated from some given theoretical framework, be it effective quark models, lattice QCD or within the framework of the Bethe-Salpeter equation applied to QCD. In the framework of QCDF the weak effective Hamiltonian of the Standard Model \cite{Beneke:1999br, Antonelli:2009ws} allows to express the amplitude of a heavy-meson decay into two light mesons as a sum of effective QCD Wilson coefficients (short distance or hard contributions) multiplied by a product 
of two currents (long-distance or soft terms), one of which generates a final meson from the vacuum and is nothing else but the weak decay constant, while the other describes the transition from a heavy meson into the second light meson~\cite{El-Bennich:2009gbu}. One of the  nonperturbative ingredients in these calculations are precisely the LCDA of the initial heavy meson and the light(er) daughter meson. 
Obtaining the broader, non-asymptotic LCDA for the $D$ mesons will allow for much needed improvement in this field.

Herein we re-appreciate earlier work on heavy-light mesons and quarkonia~\cite{Rojas:2014aka,El-Bennich:2016qmb,Mojica:2017tvh,Bedolla:2015mpa,Raya:2017ggu,
Fischer:2014cfa,Hilger:2017jti} within a continuum approach to the hadron bound-state problem in order to compute the LCDA. 
The crucial difference in the present approach is the  flavor-dependence of the interaction in the ladder truncation of the Bethe-Salpeter kernel, 
as we effectively take into account that the  quark-gluon vertex dressing has a different impact for a light quark than for a charm or bottom quark. In general, $D$ and 
$B$ mesons are of particular interest, as they offer a rich laboratory to study two limiting mass-scale sectors of QCD with associated emergent approximate symmetries: 
chiral symmetry in the sector of light quarks, where  $m_q \ll \Lambda_\mathrm{QCD}$, and heavy quark symmetry for masses $m_q \gg \Lambda_\mathrm{QCD}$
\cite{ElBennich:2012tp,Manohar:2000dt}.  

\section{Bethe-Salpeter Amplitudes and Dressed-Quark Propagators}

The calculation of a meson's distribution amplitude requires the knowledge of its wave function, which is obtained from the solution of a continuum bound state problem described by the homogeneous Bethe-Salpeter equation (BSE) in a symmetry-preserving truncation~\cite{Serna:2020txe}.

 For a given meson made up of a quark of flavor $f$ and an antiquark of flavor $g$, the BSE reads, 
\begin{equation}
\label{BSE}
  \Gamma^{fg}_M (k,P) \!=\! \int^\Lambda\! \! \!\frac{d^4q}{(2\pi)^4 } \, K_{fg} (k,q;P)  \chi_M (q_\eta, q_{\bar \eta})  ,
\end{equation}
where $K_{f g}(q, k ; P)$ is the fully-amputated quark-antiquark scattering kernel and the Dirac- and color-matrix indices are implicit. The term $ \chi_M (q_\eta, q_{\bar \eta}) =S_f (q_\eta) \Gamma^{fg}_M (q,P) S_g(q_{\bar \eta} )$ defines the Bethe-Salpeter wave function of the meson, where $S_f(q_\eta)$ and $S_g(q_{\bar\eta})$ are the dressed quark and antiquark propagators with the shorthand for the quark momenta, $q_\eta = q+ \eta P$ and $q_{\bar \eta} = q - \bar \eta P$, with the momentum-fraction  parameters, $\eta +\bar \eta = 1$, $\eta \in [0,1]$.  $ \Gamma^{fg}_M (q,P) $ is the Bethe-Salpeter amplitude (BSA) of the meson. 

We observe that iEq.~\eqref{BSE} requires the dressed-quark propagators which can be obtained from the solution of the quark Dyson-Schwinger equation, namely,
 \begin{equation}
S^{-1}_f (p)  =  \, Z_2^f  \left (i\, \gamma \cdot  p + m^{\mathrm{bm}}_f \right ) 
                     +  \, Z_1^f g^2 \!\! \int^\Lambda  \hspace*{-1.5mm} \frac{d^4k}{(2\pi)^4}  \, D^{ab}_{\mu\nu} (q) \frac{\lambda^a}{2} \gamma_\mu S_f(k) \Gamma^b_{\nu,f}  (k,p) \, ,
\label{QuarkDSE}
\end{equation}
where $m^\textrm{bm}_f$ is the bare current-quark mass, $Z_1^f(\mu,\Lambda)$ and $Z_2^f(\mu,\Lambda)$ are the vertex and  wave-function renormalization 
constants at the renormalization point $\mu$, respectively. The integral is over the dressed-quark propagator $S_f(k)$, the dressed-gluon propagator $D_{\mu\nu}(q)$ 
with momentum $q=k-p$ and the quark-gluon vertex, $\Gamma^a_\mu (k,p) = \frac{1}{2}\,\lambda^a \Gamma_\mu (k,p) $, with the SU(3) color matrices $\lambda^a$ 
in the fundamental  representation; $\Lambda$ is a Poincar\'e-invariant regularization scale, chosen such that $\Lambda \gg \mu$.

The general form of the BSA of a pseudoscalar meson is written in terms of four scalar functions and Lorentz covariants and is given by,
\begin{equation}
   \Gamma^{fg}_M (k,P)   =   \gamma_5    \Big [ i E^{fg}_M (k,P) +  \gamma \cdot P\,  F^{fg}_M (k,P)   
+   \gamma \cdot k\;  k \cdot P \,    G^{fg}_M (k,P)  +   \sigma_{\mu\nu} k_\mu P_\nu \, H^{fg}_M (k,P) \Big ] \, ,
 \label{PS-BSA}                                      
\end{equation}    
while the quark propagator is written in terms of two scalar functions
\begin{equation}
\label{DEsol}
   S_f (p)  \ =  \ -i \gamma \cdot p \, \sigma_{\rm v}^f ( p^2 ) + \sigma_{\rm s}^f ( p^2 )
                        =   \, Z_f (p^2 ) / \left [ i \gamma \cdot p + M_f ( p^2 ) \right  ] \ .
\end{equation}
The scalar functions are $\sigma_{\rm s}^f ( p^2 )$ and $\sigma_{\rm v}^f ( p^2 )$, respectively, whereas $Z_f(p^2)$ defines the quark's wave function and 
$M_f (p^2) = B_f (p^2)/A_f (p^2)$ is the running mass function. In a subtractive renormalization scheme the two renormalization conditions,
\begin{align}
   Z_f (p^2) & =   \left. 1/A_f (p^2)  \right |_{p^2 = \mu^2} = 1 \  ,
\label{EQ:Amu_ren}   \\
   S^{-1}_f (p)  &     \left.  \right |_{p^2=\mu^2}   =  \,  i \gamma\cdot p \ + m_f(\mu ) \ ,
 \label{massmu_ren}
\end{align}
are imposed, where $m_f(\mu )$ is the renormalized current-quark mass related to the bare mass by,
\begin{equation}
\label{mzeta} 
   Z_4^f  (\mu,\Lambda )\, m_f(\mu)  = Z_2^f  (\mu,\Lambda ) \, m_f^{\rm bm} (\Lambda) \  ,
\end{equation}
 and $Z_4^f(\mu,\Lambda )$ is the renormalization constant associated with the mass term in the QCD Lagrangian.
 
The solution of Eq.~\eqref{BSE} yields the mass and the Bethe-Salpeter wave function of the meson, which can be projected on the light-front to extract the distribution amplitudes. It also allows to 
obtain the leptonic decay constant, which directly tests the wave function normalization of the meson:
\begin{equation}
\label{fdecay} 
  f_M  P_\mu = \frac{N_c Z_2}{\sqrt{2} }\int^\Lambda\!  \frac{d^4k}{(2\pi)^4} \,\operatorname{Tr}_D \left [ \gamma_5\gamma_\mu\,  \chi_M (k_\eta,k_{\bar \eta}) \right ] \, .
\end{equation}

The rainbow-ladder (RL) truncation of the integral equation~\eqref{QuarkDSE} and of the BSE kernel has proven to be a robust and successful symmetry-preserving approximation
of the full tower of equations in QCD, and allows for the description of light ground-state mesons in the isospin-nonzero pseudoscalar and vector channels, as well as of the 
$N$, $N^*$ and $\Delta$ baryons~\cite{Maris:2003vk,Mader:2011zf,Bashir:2012fs,El-Bennich:2016qmb,Eichmann:2016yit,Qin:2018dqp,Yin:2019bxe}. The RL truncation is 
realized by limiting the fully dressed quark gluon vertex to the perturbative vertex:  $\Gamma_{\nu,f} \to \gamma_\nu$. We proceed the same way, however we introduce 
a  flavor-dependent ansatz  for the interaction model as follows:
\begin{equation}
 Z_1^f g^2  D_{\mu\nu} (q) \Gamma_{\nu, f} (k, p) = \big ( Z^f_2 \big)^{\!2}\, \mathcal{G}_{fg} (q^2) D_{\mu\nu}^\mathrm{free} (q) \frac{\lambda^a }{2} \gamma_\nu\, ,
\label{RLtrunc}
\end{equation}
\begin{equation}
  \label{RLkernel}
    K_{fg}(k,q;P)   = -  \mathcal{Z}_2^2  \  \frac{\mathcal{G}_{fg}  (l^2)}{l^2 } \frac{\lambda^a }{2} \gamma_\nu \frac{\lambda^a }{2} \gamma_\nu \ .
\end{equation}
Namely, in Eqs.~\eqref{RLtrunc} and \eqref{RLkernel} we combine the wave-function renormalization constant of both quarks, $ \mathcal{Z}_2 (\mu,\Lambda) =\sqrt{ Z_2^fZ_2^g}$, and 
use a flavored interaction ansatz~\cite{Serna:2020txe},
\begin{equation}
    \frac{ \mathcal{G}_{fg}  (l^2) }{ l^2 }= \mathcal{G}_{fg}^\mathrm{ IR}(l^2) +  4\pi \tilde\alpha_\mathrm{PT}(l^2) ,
\end{equation}
which leads to a different treatment of the light and heavy quarks and where the averaged interaction in the low-momentum domain is described by,
\begin{equation}
\label{BSEflavor}
 \mathcal{G}_{fg}^\mathrm{IR} (l^2)  =   \frac{8\pi^2}{(\omega_f\omega_g)^2} \sqrt{D_f\,D_g }\,e^{-l^2/(\omega_f\omega_g)} \ ,
\end{equation}
while the perturbative part is the usual one, 
\begin{align}
\label{DSEflavor} 
  4\pi \tilde\alpha_\mathrm{PT}(q^2) &=  \frac{8\pi^2  \gamma_m \mathcal{F}(q^2)}{ \ln \left [  \tau +\left (1 + q^2/\Lambda^2_\textrm{\tiny QCD} \right )^{\!2} \right ] }  \ , 
\end{align}
with $\gamma_m=12/(33-2N_f)$ being the anomalous dimension and $N_f $ the active flavor number, $\Lambda_\textrm{\tiny QCD}=0.234$~GeV, $\tau=e^2-1$, $\mathcal{F}(q^2)=[1-\exp(-q^2/4m^2_t)]/q^2$ and $m_t=0.5$~GeV.  The set of parameters used for the IR part are reported in Ref.~\cite{Serna:2020txe}.

The last step to compute meson properties implies the knowledge of the quark propagator for complex momenta,  
\begin{equation}
   S_f (q_\eta)  =   -i \gamma \cdot q_\eta \, \sigma_{\rm v}^f (q_\eta^2 )+\sigma_{\rm s}^f (q_\eta^2 )\,,
\end{equation}
and likewise for $ S_f (q_{\bar\eta } )$, as in Euclidean space the arguments $q_\eta^2$ and $q_{\bar\eta}^2$ define parabolas on the complex plane, 
\begin{eqnarray}
   q^2_\eta  & = &  q^2  - \eta^2 m^2_M  + 2 i \eta \, m_M  | q | z_q \  , \\  
   q^2_{\bar\eta}  & = &  q^2  - {\bar\eta}^2 m^2_M  - 2 i \bar \eta \, m_M  | q | z_q \  ,
\end{eqnarray}
where $z_q = q\cdot P /|q||P|$, $-1 \leq z \leq +1$, is an angle. 
The meson's masses are obtained from the solutions of the eigenvalue problem of the BSE. They are listed along with the leptonic decay constants of ground-state pseudoscalar mesons
 in Table~\ref{tab:psproperties}, from which it is clear that they are in very good agreement with experimental data, when available, or lattice-QCD results otherwise.

\begin{table}[t!]
\centering
\begin{tabular}{c|c|c|c||c|c|c}
\hline \hline
      & $m_{P}$ &$M^\mathrm{exp}_P$   &  $\epsilon_{m_P}$ [\%]    &   $f_P $  &  $f^\mathrm{exp/lQCD}_P $  &  $\epsilon_{f_P} $ [\%]  \\ [0.5mm]  
           \hline
 $\pi (u \bar d)$&0.140&0.138&1.45 &   $0.094^{+0.001}_{-0.001}$ & 0.092(1)  & 2.17  \\ \hline
 $K (u \bar s)$  &0.494&0.494&0& $0.110^{+0.001}_{-0.001}$  &0.110(2)  &  0 \\\hline 
 $D (c \bar d) $  & $1.867^{+0.008}_{-0.004} $ &1.864 &0.11 &  $0.144^{+0.001}_{-0.001}$  &0.150 (0.5) & 4.00   \\\hline 
 $D_s(c \bar s) $ &  $2.015^{+0.021}_{-0.018}$ &1.968 & 2.39 & $0.179^{+0.004}_{-0.003}$  & 0.177(0.4) &1.13  \\\hline
 $\eta_c(c \bar c) $  & $3.012^{+0.003}_{-0.039}$  &2.984&0.94 &  $0.270^{+0.002}_{-0.005}$ &0.279(17) & 3.23 \\\hline
 $\eta_b(b \bar b) $  & $9.392^{+0.005}_{-0.004}$  &9.398 & 0.06 & $0.491^{+0.009}_{-0.009}$ &  0.472(4) & 4.03   \\\hline
 $B(u\bar b)$  &    $5.277^{+0.008}_{-0.005} $  & 5.279 & 0.04 & $0.132^{+0.004}_{-0.002}$  &0.134(1) &4.35   \\\hline
 $B_s(s\bar b)$  &  $5.383^{+0.037}_{-0.039}$   &5.367 &  0.30 &  $0.128^{+0.002}_{-0.003}$   &0.162(1) & 20.5 \\\hline
 $B_c(c\bar b)$ & $6.282^{+0.020}_{-0.024}$   & 6.274 & 0.13 &  $0.280^{+0.005}_{-0.002}$ & 0.302(2) & 10.17 \\
\hline \hline
\end{tabular}

\caption{\label{tab:psproperties} Masses and decay constants [in GeV] of pseudoscalar mesons \cite{Serna:2020txe}. Experimental masses and leptonic decay constants
are taken from the Particle Data Group~\cite{PDG}. The $D$ and $D_s$ decay constants are FLAC 2019 averages~\cite{Aoki:2019cca}
and $f_{\eta_c}$  is from Ref.~\cite{McNeile:2012qf}.  The relative deviations from experimental values are  given by $\epsilon^v_r  =100\% \, | v^\textrm{exp.} 
- v^\textrm{th.} | / v^\textrm{exp.}$. }
\vspace*{-5mm}
\end{table}

\begin{figure}[b!] 
\centering
  \includegraphics[scale=0.57,angle=0]{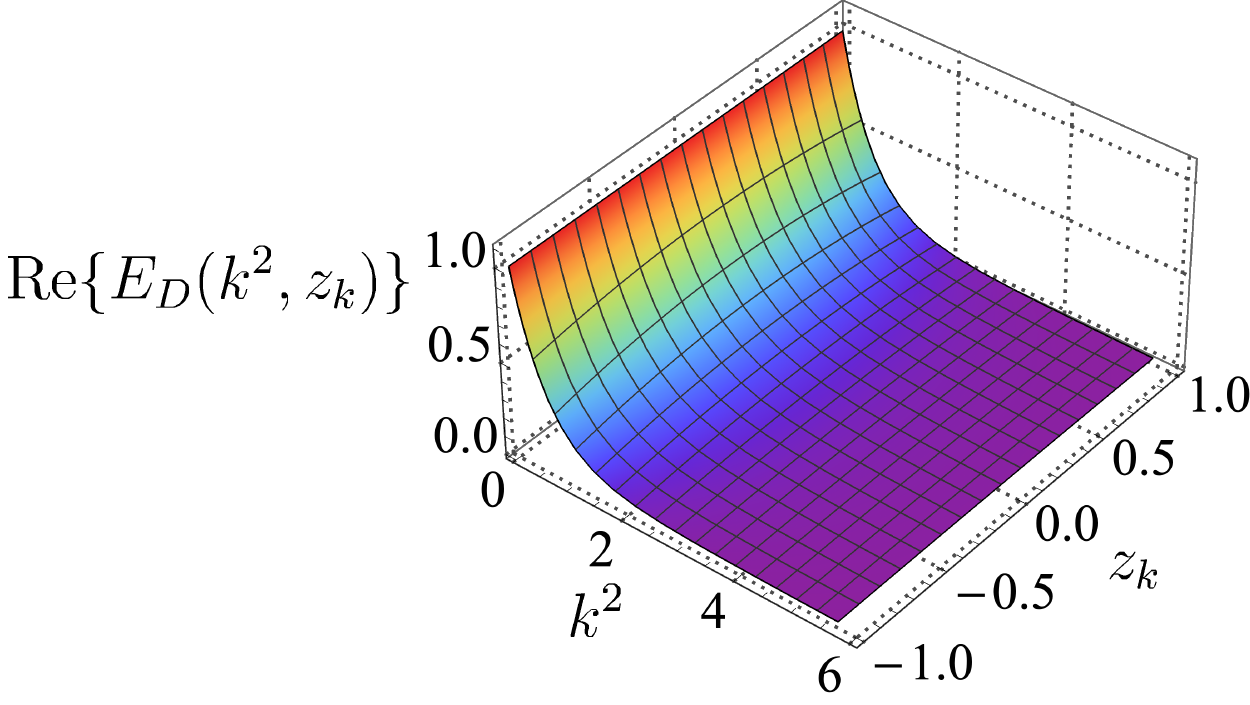} \hfill
  \includegraphics[scale=0.57,angle=0]{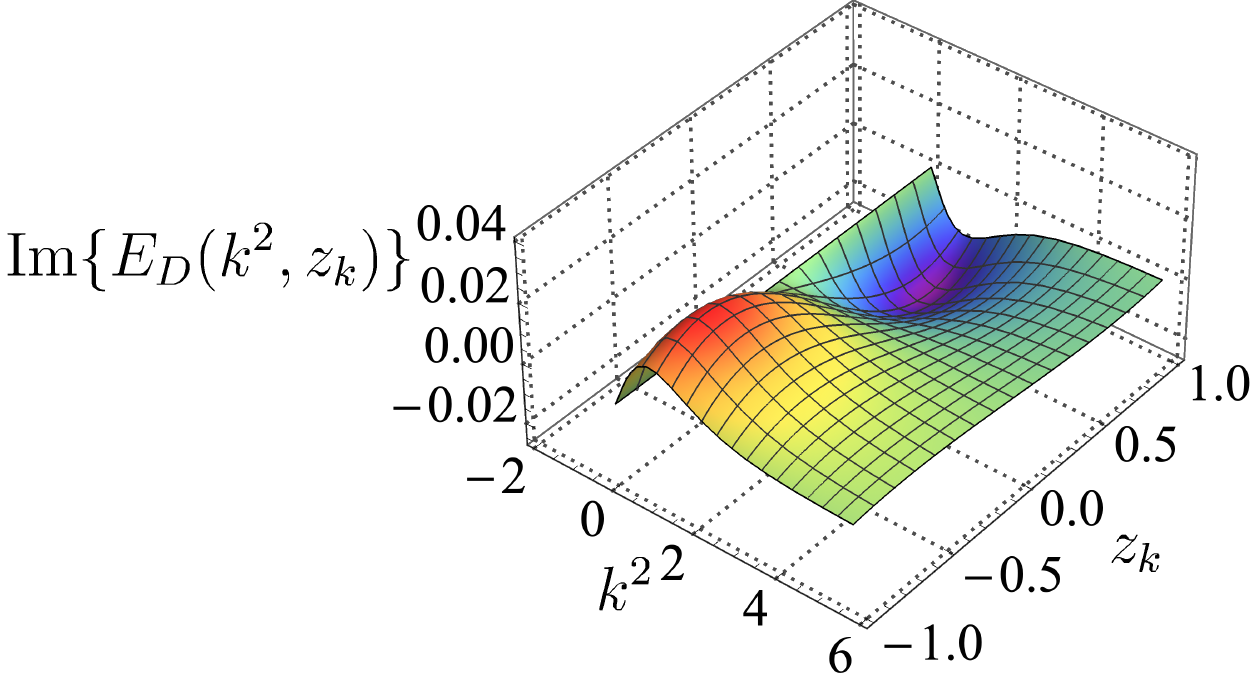}   \\ \vspace*{7mm}
   \includegraphics[scale=0.57,angle=0]{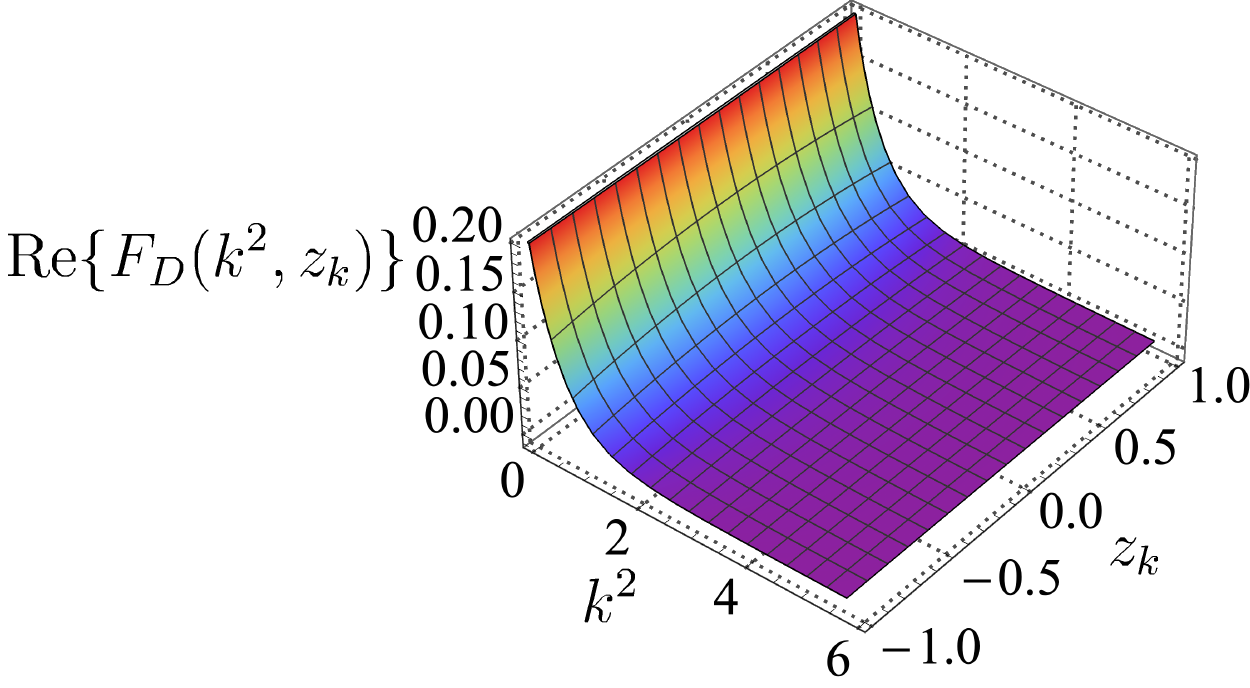} \hfill
  \includegraphics[scale=0.57,angle=0]{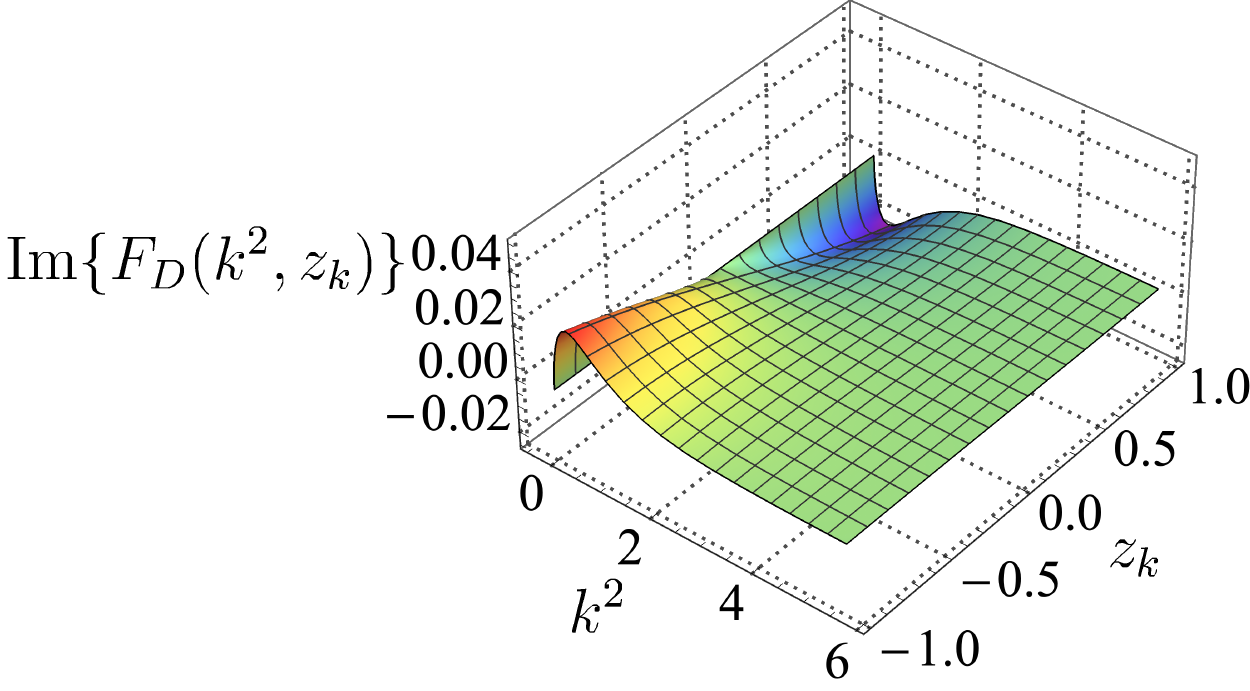} 
  
   \caption{\label{fig:E-F} Dominant Bethe-Salpeter amplitudes of the $D(c\bar u)$ meson.}
\end{figure}

Along with the bound-state masses, we also obtain the Bethe-Salpeter amplitudes and in Figure~\ref{fig:E-F}  we plot the dominant scalar functions of the $D(c\bar u)$ meson.

\section{Distribution Amplitudes}
Consider a pseudoscalar meson with mass $m_M$ constituted from a quark of flavor $f$ and an antiquark of flavor $g$; then, one may define a distribution amplitude for this system as the 
following light-front projection of the meson Poincar\'e-covariant Bethe-Salpeter wave function \cite{Chang:2013pq,Serna:2020txe}:
\begin{equation}
  f_M \phi_M (x , \mu) =  \; \frac{ \mathcal{Z}_2 N_c}{\sqrt{2}} \;\mathrm{Tr}_D\! \int^{\Lambda}  \!\!  \frac{d^4k}{(2\pi)^4} \, \delta  (n \cdot k_\eta - x n\cdot P )
 \gamma_5 \, \gamma \cdot n \ \chi_M (k_\eta,k_{\bar \eta})\, ,
\label{BSA-LCDA}  
\end{equation}
where $f_M$ is the weak decay constant of the pseudoscalar meson, $n$ is an auxiliary light-like four-vector with $n^2 =0$, $x$ is the momentum fraction of the quark and 
$n \cdot P = - m_M$ in the rest-frame of the meson. 

Our DSE and BSE framework is formulated in Euclidean metric, analogous to 
lattice QCD, and therefore a direct calculation of a light-like correlation function, such
as $\phi_M (x , \mu)$, is difficult, although a potential solution within this approach
is provided by the Nakanishi representation \cite{Chang:2013pq}. In this work we use the Mellin moments
method to determine the leading-twist LCDA. We define the Mellin moment $m$ of a LCDA by, 
\begin{equation}
   \left\langle x^{m}\right \rangle = \int_{0}^{1} dx\,  x^{m} \phi_M(x,\mu)  \ ,
\label{momentadef}   
\end{equation} 
using Eq.~\eqref{BSA-LCDA} and then reconstruct the LCDA from these moments. In particular, the zeroth moment serves to normalize the distribution amplitude
and we choose,
\begin{equation}
  \langle x^{0}  \rangle =  \int_{0}^{1} d x \, \phi_M (x,\mu )  = 1 \ .
\label{zeromomentum}  
\end{equation}
With this, one may apply the integral in Eq.~\eqref{momentadef}  to both sides of Eq.~\eqref{BSA-LCDA} which, employing the property of the 
Dirac function $\int_{0}^{1} d x\, x^{m} \delta(a-x b) = \frac{a^{m}}{b^{m+1}} \,\theta(b-a) $, leads to the integral, 
\begin{eqnarray}
  \left  \langle x^{m}\right \rangle    =    \frac{\mathcal{Z}_2N_c}{\sqrt{2}f_M} \; \mathrm{Tr}_D \! \int^{\Lambda}  \!\! \frac{d^4k}{(2\pi)^4} 
  \frac{ ( n\cdot k_\eta )^m}{ ( n\cdot P )^{m+1}  }   
\   \gamma_5 \, \gamma \cdot n \; \chi_M (k_\eta,k_{\bar \eta})\, .
  \label{moment-BSA}          
\end{eqnarray}
The moments $\left\langle x^{m}\right \rangle$ and therefore the reconstructed distribution amplitudes are valid at a given scale at which the
BSA was calculated. All our results are given for a fixed scale: $\mu = 2$~GeV.  Then, in order to compute 
the moments $\left\langle x^{m}\right \rangle$, we first parametrize our solutions for the quark propagators by using a complex conjugate pole (\texttt{ccp}) parametrization written as the sum, 
\begin{equation}
\label{ccp}
 S_f  (q )  =  \sum^N_{k=1}  \left [   \frac{z_k^f}  {i \gamma\cdot q  +  m_k^f  }  +   \frac{\big (z_k^f\big )^{\!*} }  {i \gamma\cdot q  +  \big (m_k^f\big )^{\!*}  } \right  ] \ ,
\end{equation}
where $m_k^f$ and $z_k^f$ are complex numbers~\cite{Bhagwat:2002tx}. These parameters are fitted to the DSE solution~\eqref{DEsol} for $N=2$ on the real space-like axis $p^2\,\in[0,\infty)$, 
and the thus obtained \texttt{ccp} representation is then analytically extended to complex momenta. 
On the other hand, for the case of the meson's BSAs we use a Nakanishi representation \cite{Chang:2013pq} (perturbation theory integral representation) of the scalar amplitudes 
$\mathcal{ F}_i= E_M$, $ F_M$, $G_M$, $H_M $:
\begin{equation}
 \label{BSAparKD}
 \hspace*{-2mm}
   \mathcal{F}_i (k,P)=\sum^{2}_{\sigma=0}  \,  \int^1_{-1}  \!dz\,  \rho_{\nu_\sigma}(z) 
       \frac{\mathcal{U}_\sigma\Lambda^{2n_\sigma}_{\mathcal{F}_i}}{(k^2+z\, k\cdot P + \Lambda^2_{\mathcal{F}_i})^{n_\sigma}} \, ,
\end{equation}
using $\mathcal{U}_0 = U_0-U_1-U_2$, $\mathcal{U}_1 = U_1$ and $\mathcal{U}_2 = U_2$, and the spectral density is given by,
\begin{equation}
    \rho_{\nu_\sigma} = \frac{\Gamma\left(\nu_\sigma+3/2\right)}{\sqrt{\pi}\,\Gamma\left(\nu_\sigma+1\right)}(1-z^2)^{\nu_\sigma}.
 \label{rhropion}   
\end{equation}
The scalar amplitude $H(k,P)$ is negligibly small, has little impact, and is thus neglected. The advantage of using such representations is that it allows us to write the moments in 
Eq.~\eqref{moment-BSA} in terms of Feynman integrals. Details of the set of parameters obtained for the \texttt{ccp} representation for the quark propagators and the Nakanishi 
representation for BSAs can be found in Ref.~\cite{Serna:2020txe}.


\section{Calculated Distribution Amplitudes}
\begin{figure}[t!] 
\centering
  \vspace*{-1cm}
  \includegraphics[scale=0.95,angle=0]{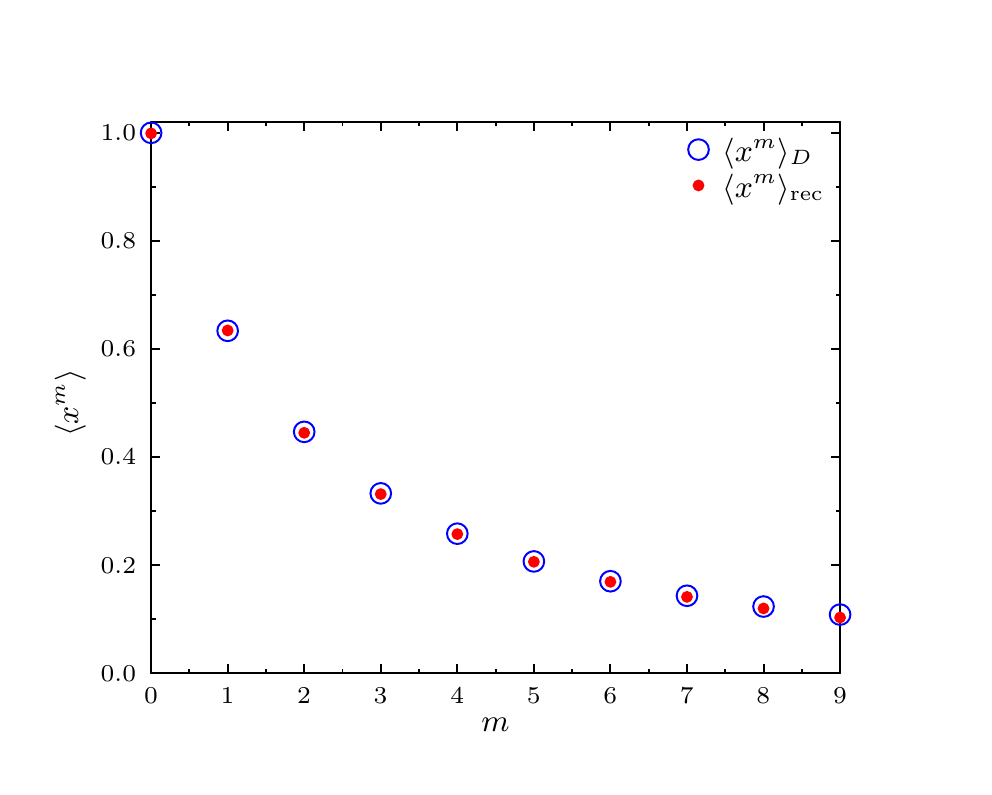}  
  \vspace*{-4mm}
   \caption{\label{fig:moments-D} Values of the moments $\langle x^m \rangle_D$ and $\langle x^m \rangle_{\rm rec}$ as a function of the power $m$.. }
\end{figure}

In order to reconstruct the distribution amplitudes $\phi_M(x,\mu)$ from the light-meson moments, we 
write them in terms of Gegenbauer polynomials, $C_n^\alpha (2x-1)$, of order $\alpha$ which
form a complete orthonormal set on $x\in [0,1]$ with respect to the measure $[x(1-x)]^{\alpha-1/2}$.
In general, for a flavored meson such as the  kaon, we reconstruct its LCDA with the parity decomposition,
\begin{equation}
\label{phi-recon}
   \phi_M^\mathrm{rec.}(x,\mu) = \phi^E_M (x,\mu)  + \phi^O_M (x,\mu) \ ,
\end{equation}
where we employ one and two Gegenbauer polynomials, respectively, in the even and odd components, 
\begin{subequations}
\begin{eqnarray} 
  \label{evenpda}  
  \phi^E_{M}(x,\mu )    & =  &  \frac{\Gamma(2\alpha+1)}{[\Gamma(\alpha+1/2)]^2} \, [x \bar x]^{\alpha-\frac{1}{2}}\big [1+a_2 C^\alpha_2(2x-1) \big ] ,  \\ [0.2true cm]   
  \phi^O_{M}(x,\mu )   &  =  &  \frac{\Gamma(2\beta+1)}{[\Gamma(\beta+1/2)]^2}  \, [x \bar x]^{\beta-\frac{1}{2}} \big [ b_1 C^\beta_1(2x-1) 
                                    +     b_3 C^\beta_3 (2x-1)  \big ] \ .
 \label{oddpda}                                        
\end{eqnarray}
\end{subequations}
These even and odd parts of the distribution amplitudes are then determined independently by separately minimizing,
\begin{align}
  \epsilon_E  (\alpha, a_2) & =  \sum_{m=2,4,..., 2m_\textrm{max}}  \left | \frac{\langle \xi^m\rangle^E_\mathrm{rec.}}{\langle \xi^m\rangle_M } -1 \right| \ , \\
  \epsilon_O (\beta,b_1,b_3) & =  \sum_{m=1,3,..., 2m_\textrm{max}-1}   \left | \frac{\langle \xi^m\rangle^O_\mathrm{rec.}}{\langle \xi^m\rangle_M } -1 \right | \ ,
\end{align}
with 
\begin{equation}
\label{kaonmoments}
  \langle  \xi^m  \rangle_M  \  =    \int_{0}^{1} dx \, (2 x-1)^m  \phi_K (x, \mu ) \ ,\,\,\,  \xi  =x-(1-x) =  2 x-1 \ ,
\end{equation}
and the reconstructed moments $\langle \xi^m\rangle^{E,O}_\mathrm{rec.}$ are obtained with the distribution amplitudes in Eqs.~\eqref{evenpda} and \eqref{oddpda}. For equal valence-quark mesons like the pion, 
which are eigenstates of charge conjugation, the odd contributions in Eq.~\eqref{phi-recon} vanish. We remind that in the asymptotic limit the pion LCDA tends to~\cite{Lepage:1979zb}: 
\begin{equation}
      \phi_\pi (x,\mu)\  \stackrel{\mu \to \infty }{\longrightarrow}  \  \phi_{\mathrm{asy}} = 6x (1- x) =  6x \bar x \ .
\end{equation}
The heavy mesons are treated similarly, yet we employ a different functional form for $\phi_{H}^\mathrm{rec.}$ given by~\cite{Ding:2015rkn}, 
\begin{equation}
\label{phi-heavy}
   \phi_H^\mathrm{rec.} (x,\mu) = \mathcal{N}(\alpha,\beta) \,  4x\bar x\,e^{4\,\alpha x\bar x+ \beta (x-\bar x)} \ .
\end{equation}
We take the opportunity to correct an error in the expression for the normalization $\mathcal{N}(\alpha,\beta)$ in Eq.~(58) of Ref.~\cite{Serna:2020txe}. With the definition of the error function, 
$\operatorname{Erf}(x)  = \frac{2}{\sqrt{\pi}}\int^x_0 dt\, e^{t^2}$, this becomes:
\begin{align}
   \mathcal{N}(\alpha, \beta )   = & \ 16\,\alpha^{5/2} \Bigg [ 4\sqrt{\alpha}\, \left ( \beta \sinh(\beta)+2a\cosh(\beta) \right )  \nonumber \\ 
           + &\sqrt{\pi}  \ e^{\alpha+\frac{\beta^2}{4\alpha}} \left ( -2\alpha+4\alpha^2 - \beta^2 \right ) 
  \left \{ \operatorname{Erf}\left(\frac{2\alpha-\beta}{2\sqrt{\alpha}}\right) + \operatorname{Erf} \left( \frac{2\alpha+\beta}{2\sqrt{\alpha}} \right ) \right \}  \Bigg ]^{-1} \  .
\end{align}
The reason for this choice is that the Gegenbauer procedure sketched above is appropriate for broader and concave amplitudes, whereas a 
distribution amplitude with a convex-concave behavior reminiscent of the $\delta$-function in the infinite-mass limit is more appropriately
described by Eq.~\eqref{phi-heavy}. 
 Moreover, using the separation in even and odd components with Eqs.~\eqref{evenpda} 
and \eqref{oddpda} for the $D$ and $B$ mesons requires the computation of a large number of Mellin moments to fix their coefficients. The larger moments 
suffer numerical instabilities in that case and we therefore prefer the representation in Eq.~\eqref{phi-heavy}. We thus reconstruct the LCDA by minimizing, 
\begin{equation}
   \epsilon  (\alpha, \beta) = \sum^{m_\textrm{max}}_{m=1}  \left | \frac{\langle x^m \rangle_\textrm{rec.}}{\langle x^m\rangle_H } -1 \right | \ ,
\end{equation}
with $\langle x^m \rangle _\textrm{rec.}$ calculated as described before and making use of Eq.~\eqref{phi-heavy}. In Figure~\ref{fig:moments-D} we present the results obtained in our 
reconstruction process for the $D$-meson with $\alpha =  0.038\pm0.005$ and $\beta  = 1.431\pm 0.085$ \cite{Serna:2020txe}. 
\begin{figure}[htb] 
\centering
  \includegraphics[scale=0.74,angle=0]{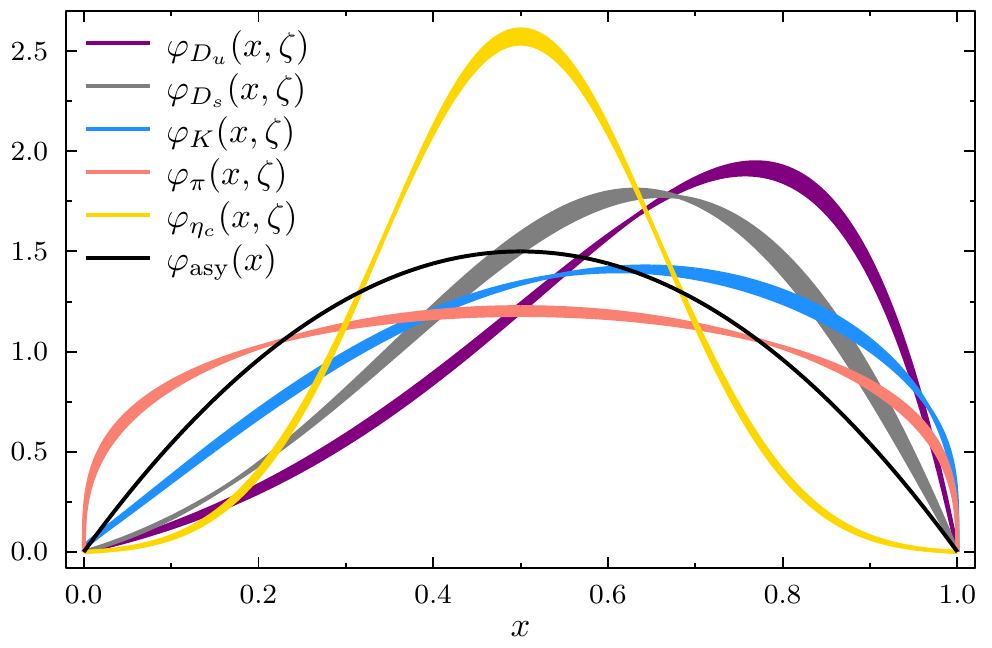}  
    \includegraphics[scale=0.74,angle=0]{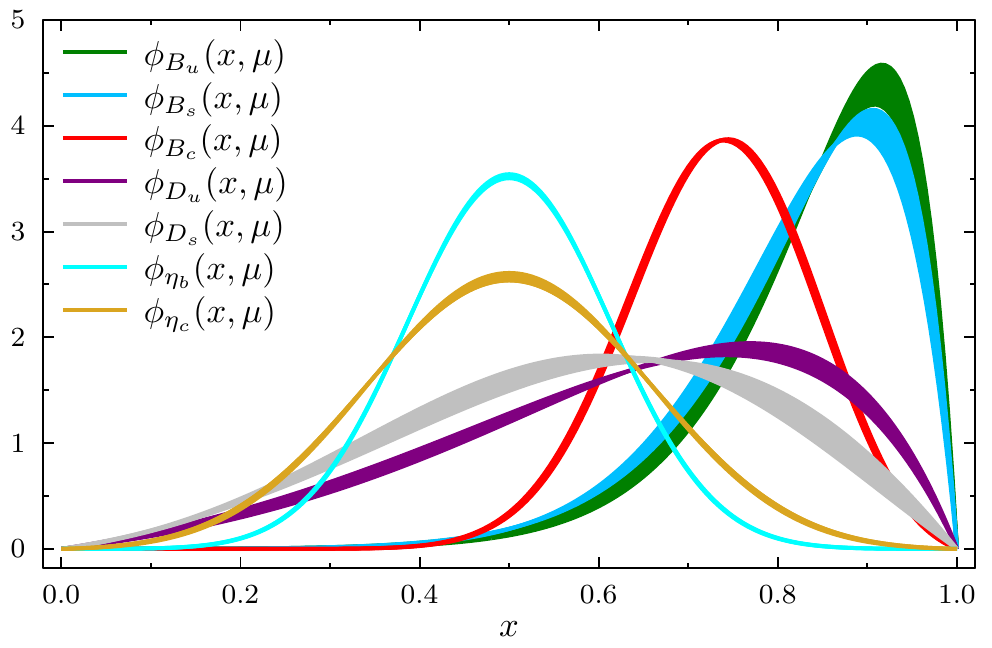}  
   \caption{\label{fig:PDAs}Distribution amplitudes on the light front at a renormalization point $\mu = 2$~GeV \cite{Serna:2020txe}.  
   \textbf{Left panel}: Comparison of the light-meson distribution amplitudes $\phi_\pi (x, \mu )$, $\phi_K(x,\mu )$  with $\phi_{D_u} (x,\mu)$,  $\phi_{D_s} (x,\mu)$,  $\phi_{\eta_c} (x,\mu)$, and $\phi_\mathrm{asy} (x)=6x\bar x$ is the asymptotic LCDA. 
   \textbf{Right panel}:  Comparison of the LCDAs of the charmonium, bottonium and the different $D$ and $B$ mesons. Seer Ref.~\cite{Serna:2020txe} for the origin of the error bands.  }
\end{figure}

In the left panel of Figure~\ref{fig:PDAs} we observe that $\phi_\pi(x,\mu)$ is a \emph{concave\/} and symmetric function of $x$, much broader than the asymptotic limit $\phi_\mathrm{asy} (x)$ as a consequence 
of dynamical chiral symmetry breaking (DCSB). The symmetrical shape of the pion's LCDA is precisely due to the fact that it is made of two quarks of the same flavor, each carrying the same fraction of the bound state's total momentum on the light front. 
On the other hand,  $\phi_K(x,\mu)$ turns out to be also concave and wider than $\phi_\mathrm{asy}(x)$, but with a pronounced asymmetry and its maximum is located at 
$x=0.61$. This is a clear signal of flavor $SU(3)$ symmetry breaking and demonstrates that the heaviest valence quark inside the kaon carries a greater amount of meson momentum. From the 
DSE solutions we also find the ratio $M^E_u/M^E_s=0.73$, where $M^E_f$ is the Euclidean constituent quark mass defined by: $M^E_f = \{p^2|p^2=M^2(p^2)\} $. With respect to the charmed mesons, $D_u$ and $D_s$, we note that their LCDA are now convex-concave as a function of $x$  and that the distributions are even more asymmetric than that of the kaon. Here, the heavier charm quark carries most of the meson's momenta. We also note that $\phi_{D_u}(x,\mu)$ is slightly more asymmetric  than $\phi_{D_s}(x,\mu)$, which is primarily due to the mass difference between light and strange quarks for which we obtain the ratio of constituent quarks: $M^E_u/M^E_c=0.30$ and $M^E_s/M^E_c=0.42$. Conversely, the $\eta_c$ consists of two equal quarks and is evenly distributed about $x=0.5$, however it is narrower than the asymptotic limit. 

Moving our attention to the heaviest mesons, we present a comparison between the results for the bottom and the charmed mesons in the right panel of Figure~\ref{fig:PDAs}. In this case, the 
$B$-mesons are characterized by a highly asymmetric momentum distribution and we see that the $b$-quark inside the $B_u$ and $B_s$ carries almost all of  the momentum. Their maxima are located at 
$x=0.92$ and $x=0.90$, respectively, while for $\phi_{D_u}(x,\mu)$ and $\phi_{D_s}(x,\mu)$ the peaks are located at 
$x=0.76$ and $x=0.63$. The $B_c$-meson is made of two heavy quarks, though one of them is heavier, and we note that its LCDA 
is less shifted towards $x >0.5$ with a peak at $x=0.74$. In this case we find:  $M^E_u/M^E_b=0.10$, $M^E_s/M^E_b=0.12$ and  $M^E_c/M^E_b=0.32$. Finally, we observe 
that the LCDA of the  $\eta_b$ is much narrower than that of the $\eta_c$.


\section{Conclusions}
Working within a modified RL truncation of the combined Dyson-Schwinger and Bethe-Salpeter equations, we computed the leading-twist light cone distribution amplitudes of heavy-light pseudoscalar mesons. 
The modified RL truncation consists in the introduction of a different flavor dependence in the heavy-quark sector of the Bethe-Salpeter ladder kernel, which allows for an improved 
calculation of the mass spectrum and leptonic decay constants of heavy-light mesons and quarkonia  at a physical pion mass.
With these results, we project the Bethe-Salpeter amplitudes of the pseudoscalar mesons on the light front and compute moments of the corresponding LCDA. 

The distribution amplitudes we obtain follow the expected pattern, i.e. the pion distribution amplitude is a concave function, much broader than the asymptotic one. The same is 
observed for the kaon which in addition is not symmetric about the midpoint $x=1/2$, a visual expression of SU(3) flavor breaking due to DCSB, and this asymmetry 
is growing with increasing mass of the heavier quark. The distribution amplitudes of the $D$ and $B$ mesons describe a \emph{convex-concave} function, whereas for 
the $\eta_c$ and $\eta_b$ the symmetric distribution amplitude is of \emph{convex-concave-convex} form which tends to a Dirac $\delta$ function in the infinite-mass 
limit.

\section{Acknowledgments}
We are grateful to the organizers for the opportunity of a "virtual presentation" and a well-organized on-line meeting. This work is supported by the Brazilian agencies FAPESP, grant no. 2018/20218-4, and CNPq, grant no. 428003/2018-4, and is part of the project "INCT-F\'isica Nuclear e Aplica\c c\~oes", no. 464898/2014-5. F.E.S. is a CAPES-PNPD postdoctoral fellow, contract no. 88882.314890/2013-01.

\end{document}